\documentclass[proceedings, preprint]{rmaa}

\usepackage{paralist}

\usepackage{psfrag,color}

\SetYear{2014}
\SetConfTitle{Proceedings of the astrorob conference}

\title{Web-based tools for the analysis of TAOS data and much more} 

\author{
  D. Ricci\altaffilmark{1},
  P.-G. Sprimont\altaffilmark{2},
  C. Ayala\altaffilmark{1},
  F. G. Ram\'on-Fox\altaffilmark{1},
  R. Michel\altaffilmark{1},
  S. Navarro\altaffilmark{1},
  S.-Y. Wang\altaffilmark{3},
  Z.-W. Zhang\altaffilmark{3},
  M. J. Lehner\altaffilmark{3},
  L. Nicastro\altaffilmark{2},
  M. Reyes-Ruiz\altaffilmark{1}
}

\altaffiltext{1}{Instituto de Astronom\'ia, UNAM, Campus Ensenada,
  22860 Ensenada, B.C., M\'exico (indy@astrosen.unam.mx).}
\altaffiltext{2}{INAF, Istituto di Astrofisica Spaziale e Fisica
  Cosmica di Bologna, via P. Gobetti, 101, 40129 Bologna, Italy.}
\altaffiltext{3}{ASIAA, 11F of Astronomy-Mathematics Building,
  AS/NTU. No.1, Sec. 4, Roosevelt Rd, Taipei 10617, Taiwan, R.O.C.}

\shortauthor{Ricci et al.}
\shorttitle{RevMexAA(SC) Demo Document}

\listofauthors{D. Ricci, P-G. Sprimont, C. Ayala, F. G. Ram\'on-Fox, R. Michel,
  S. Navarro, S.-Y. Wang, Z.-W. Zhang, M. J. Lehner, L. Nicastro \& M. Reyes-Ruiz}
\indexauthor{Ricci, D.}
\indexauthor{  Sprimont, P.-G.}
\indexauthor{  Ayala,     C.  }
\indexauthor{  Ram\'on-Fox, F. G.  }
\indexauthor{  Michel,    R.  }
\indexauthor{  Navarro,   S.  }
\indexauthor{  Wang    S.-Y. } 
\indexauthor{  Zhang   Z.-W. }
\indexauthor{  Lehner M. J. }
\indexauthor{  Nicastro,  L.  }
\indexauthor{  Reyes-Ruiz, M.  }

\abstract{ We suggest a new web-based approach for browsing and
  visualizing data produced by a network of telescopes, such as those
  of the ongoing TAOS and the forthcoming TAOS II projects.
  We propose a modern client-side technology and we present two
  examples based on two software packages developed for different
  kinds of server-side database approaches.
  In spite our examples are specific for the browsing of TAOS light
  curves, the software is coded in a way to be suitable for the use in
  several types of astronomical projects.}

\resumen{ Sugerimos un nuevo enfoque basado en las tecnolog\'ia web
  para navegar y visualizar datos producidos por una red de
  telescopios, tales como los del projecto TAOS en curso y del
  projecto TAOS II de pr\'oxima realizaci\'on.
  Proponemos una moderna tecnolog\'ia ``lado cliente'' y presentamos
  dos ejemplos basados en dos programas desarrollados para diferentes
  tipos de acceso ``lado servidor'' a la base de datos.
  El software se program\'o para fines astron\'omicos gen\'ericos, y
  fue aplicado espec\'ificamente en los ejemplos para la inspecci\'on
  de las curvas de luz de TAOS.  }

\addkeyword{Stars: Variables}
\addkeyword{Techniques: Photometric}
\addkeyword{Methods: Data Analysis}
\addkeyword{Astronomical Data Bases: Miscellaneous}
\addkeyword{Astronomical Data Bases: Virtual Observatory Tools}

\begin{document}
\maketitle



\section{General}
\label{intro}

\begin{figure*}[t!]
\centering
  \includegraphics[width=0.97\textwidth]{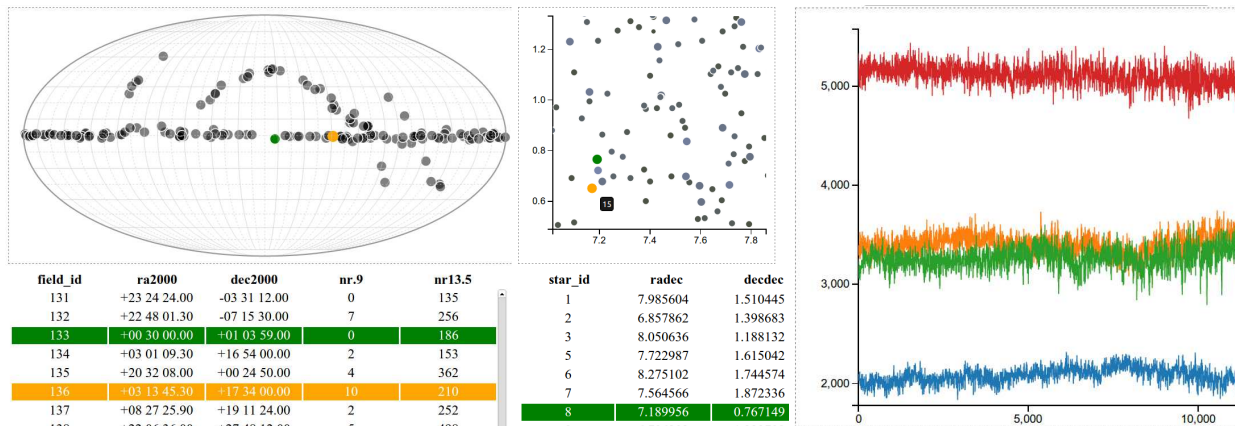}
  \caption{Example of Web2.0 approach. LEFT: page with the map of TAOS
    fields; CENTER: stars of a selected field (the axes represent
    celestial coordinates); RIGHT: light curves of a selected star
    observerd by 4 telescopes (the $x$ and $y$ axes represent points
    and counts, respectively). Colored Selections for clicked and
    mouseover elements are also shown.  The plots were populated by
    querying a \texttt{mysql} database using ``\texttt{decibel}'' (see
    the text), and rendered in \texttt{<svg>} using \texttt{d3}. }
  \label{fig1}
\end{figure*}
Modern astronomical observatories obtain every night a large amount of
information every night,
and this is even more true in the case of pre-scheduled or automatic
surveys, such as those implemented for robotic telescopes.  These
data are then typically stored in web-accessible databases.
One example of these robotic observations is represented by TAOS II
(Transneptunian Automated Occultation Survey), a project aimed at
measuring the size distribution of small objects in the Kuiper Belt
and beyond\footnote{\url{http://taos2.asiaa.sinica.edu.tw/}}.
The produced data will be automatically processed to
output light curves for more than $10\ 000$ stars at a high rate
($20\rm Hz$).  This survey will improve the previous
TAOS project
(Taiwan American Occultation Survey), which already provided a large
amount of data \citep{taos}.
How to keep trace and, when necessary, visually inspect such a huge
amount of data?
It is crucial to have a tool that can efficiently access and browse
the database and then display graphically the selected data.
Modern web 2.0 tools provide the perfect solution because they allow
to build extremely user-friendly interfaces to achieve this task in a
graphical and intuitive way, while preserving a scientific approach.
Moreover, the concept can be applied to different kind of archives.
In the Internet era, the most easy way to filter and browse a huge
amount of information is indeed a web browser.
%

\section{ Client-Side implementation}

We suggest to build this kind of new web-based astronomical
applications by using the most recent client-side {\sc html5}
technologies, following the pattern of educational examples already
realized to test the methodology \citep{grb13,inted}.  Among them
the new \verb|<canvas>| tag is noticeable for its possible use in tools
like a FITS file inspectors.  In fact it would be able to display and
manipulate PNG-tiles (like those used for ground maps), or 3-d graphs
through the use of \verb|webgl|. We also suggest the use of vector
graphics by embedding objects in \verb|<svg>| tags.
%
For a true and more advanced user-client and client-server
interaction, \verb|javascript| is commonly used; the
 \verb|jquery| library for ajax calls and the \verb|d3.js|
library\footnote{See \url{http://jquery.com/} and
  \url{http://d3js.org/}} for data-driven manipulations of the
DOM \citep{ddd} were successfully tested and recommended to this aim.

\section{ Server-Side technologies and examples}

Once the server side call is done, the data can be retrieved from a
database and pre-treated using server-side technologies. We propose
here two possible approaches.
The first is the traditional LAMP architecture: server-side Linux
machine with an Apache web server, \verb|MySQL| database management
system, and \verb|PHP| as server-side scripting language, with the
inclusion of the \verb|mysqli| module for the creation of object
oriented interfaces.  In this framework, we implemented
``\verb|decibel|'', a mysql-wrapper specifically meant for
astronomical
databases\footnote{\url{http://ross.iasfbo.inaf.it/~gloria/decibel-class/}}.
We used it to
visualize\footnote{\url{http://sadira.iasfbo.inaf.it/~indy/taos-fields/taosfields.html}}
the TAOS observed fields (see Fig.~\ref{fig1}) in the framework of a
preliminary data reduction also presented in {\bf these Proceedings (Ricci et al.)}
Eventual \verb|c++| applications for data reduction can be called in
\verb|Php| using the \verb|exec()| command as true external programs.
The second, modern approach is based on the new concept of
SSE (Server Sent Events), and implemented using \verb|Node.js|, a
solution which makes use of javascript also as server-side language to
build a web server and associated services.  
In this case, the \verb|c++| applications can be fully compiled as
\verb|Node.js| modules and integrated in ``addons''.
As an alternative to \verb|MySQL|, it is possible to use
non-relational databases such as \verb|MongoDB|, excellent to manage
tree-structured data.  The advantage of this database is that the
structure and the output is in \verb|json| format, which is light and
easy to handle in javascript.  Following this path, we contributed to
the development of the software ``\verb|sadira|'', a database system
and web GUI for astronomical data
storage\footnote{\url{http://sadira.iasfbo.inaf.it/}}, also presented
in {\bf these Proceedings (Sprimont et al.)}, in the framework of the
GLORIA (GLObal Robotic-telescopes Intelligent Array)
project\footnote{\url{http://gloria-project.eu}}. 
A test with a TAOS file containing the light curve and other data
stored in the FITS binary tables is visible in the ``{\it
  Experiments}'' section of the sadira demo site.

\section{ Development \& Applications}

The two presented solutions and the developed software
(``\verb|decibel|'' and ``\verb|sadira|'') are being developed in
parallel to provide a complementary set of tools.
The goal is to provide an easy way to inspect the database of TAOS
light curves and in general any kind of astronomical database, such
as, for example, a traditional set of images in FITS
format, or an archive of raw or reduced spectra.



\bibliography{ricci-poster-short}{}

\begin{thebibliography}
\expandafter\ifx\csname natexlab\endcsname\relax\def\natexlab#1{#1}\fi
\expandafter\ifx\csname href\endcsname\relax
  \def\href#1#2{}\fi
\expandafter\ifx\csname urllinklabel\endcsname\relax
  \def\urllinklabel{[LINK]}\fi
\expandafter\ifx\csname adsurllinklabel\endcsname\relax
  \def\adsurllinklabel{[ADS]}\fi

\bibitem[{Bostock {et~al.}(2011)Bostock, Ogievetsky, \& Heer}]{ddd}
Bostock, M., Ogievetsky, V., \& Heer, J. 2011, Visualization and Computer
  Graphics, IEEE Transactions on, 17, 2301
 \href{http://vis.stanford.edu/papers/d3}{\urllinklabel}

\bibitem[{{Ricci} \& {Nicastro}(2013)}]{grb13}
{Ricci}, D. \& {Nicastro}, L. 2013, in EAS Publications Series, Vol.~61, EAS
  Publications Series ({Castro-Tirado}, A.~J. and {Gorosabel}, J. and {Park},
  I.~H.), 263--265


\bibitem[{Ricci {et~al.}(2013)Ricci, Nicastro, \& Pio}]{inted}
Ricci, D., Nicastro, L., \& Pio, M. 2013, in INTED2013 Proceedings, 7th
  International Technology, Education and Development Conference (IATED),
  2998--3007


\bibitem[{{Zhang} {et~al.}(2013){Zhang}, {Lehner}, {Wang}, \& {et~al}.}]{taos}
{Zhang}, Z.-W., {Lehner}, M.~J., {Wang}, \& {et~al}. 2013, \aj, 146, 14


\end{thebibliography}

\end{document}